\documentclass{article}
\usepackage[final]{graphicx}
\usepackage{amsmath,amssymb}

\begin{document}

\title{Weak value as an indicator of back-action}

\author{\name{\fname{Riuji} \surname{Mochizuki}}{1}}

\author{Riuji Mochizuki\\Laboratory of Physics, Tokyo Dental College,\\ 2-9-7 Kandasurugadai, Chiyoda-ku Tokyo 101-0062, Japan}

\maketitle
\begin{abstract}%
In this study we critically examine some important papers on weak measurement and weak values.  We find some insufficiency and mistakes in these papers, and we demonstrate that the real parts of weak values provide the back-action to the post-selection, which is caused by weak measurement.  Two examples, a counterfactual statement of Hardy's paradox and experiments that determine the average trajectory of photons passing through double slits, are investigated from our view point.

PACS numbers: 03.65.Ta, 03.65.Ud, 03.65.Ca
\end{abstract}

\section{Introduction}

Since Aharonov {\it et al}.\cite{Aha2}\cite{Aha25}\cite{Aha3} developed the concepts of weak measurement and weak values, these ideas have been studied by many authors.  In weak measurement, which differs from conventional  von Neumann-type measurement\cite{von} ({\it strong} measurement in this paper), the interaction between an observed system and a probe is considered to have no effect on the observed system when its weak coupling limit is taken.   Some authors\cite{Aha3}\cite{Hosoya1}\cite{Tresser1} have even claimed that noncommuting observables can be measured simultaneously by weak measurement, and relations to Bell's inequality\cite{Bell} have also been discussed.  In addition, it has been claimed that wave functions can be directly determined by weak measurement\cite{Hof1}\cite{Lund}\cite{Lund2}.  In particular, Wiseman\cite{Wise1} defined the average velocity of photons operationally with weak measurement and demonstrated identification with Bohm's velocity\cite{Bohm}.  Kocsis {\it et al}. have developed this study and reported\cite{Kocsis} that the average trajectory of photons passing through double slits can be determined operationally by weak measurement maintaining the interference pattern.

Weak values have attracted attention because of both the values obtained by  weak measurement and their inherent physical meaning\cite{Hosoya2}.   For example, the counterfactual statements of Hardy's paradox and the three-box paradox have been interpreted with the help of weak values\cite{Hardy}\cite{Aha4}\cite{Aha6}\cite{Vaidman}, which were experimentally verified\cite{Irvine}\cite{Lundeen}\cite{Yokota}\cite{Resch} to agree with the values obtained by  the corresponding weak measurements.    Moreover,  strange weak values have been discussed by many authors\cite{Leggett}\cite{Duck}\cite{Hosoya2}\cite{Dressel}, but the conditions in which they appear have not been clarified.

Despite the strange properties of weak values, they have been interpreted as conditional probabilities or conditional expectation values by many authors.  One of their main bases is that ordinary expectation values can be described as the sum of weak values, which was demonstrated in \cite{Aha3}.  To corroborate the above statement, other authors\cite{Wise2}\cite{Dre2}\cite{Dressel}\cite{Dre3}\cite{Hof1}\cite{Hof2} have discussed this problem with {\it positive operator-valued measure} (POVM).  In the second section, we examine their discussions, and we find some insufficiency and mistakes.  Then, we demonstrate that the real parts of weak values should be interpreted as the indicator of the {\it back-action} caused by the weak measurement.  Reinvestigation of the operational process of the post-selection provides a clearer basis for the above conclusion.  In the following two sections, Hardy's paradox and the double-slit experiment are investigated from the viewpoint given in the second section.  The last section is our conclusion.

\section{Interpretation of weak values}

\subsection{Weak measurement and weak values}
First, we quickly review the relation between weak values and the values obtained by the corresponding weak measurement\cite{Aha2}\cite{Lun1}.  
The interaction Hamiltonian $\hat H_I$ between an observable $\hat A$ of the quantum system and the momentum $\hat \pi$ of the pointer is
\begin{equation}
\hat H_I\equiv g\hat A\hat \pi,\label{eq:Hamiltonian}
\end{equation}
where $g$ is the real coupling constant.   $\hat H_I$ is assumed to be constant and roughly equivalent to the total Hamiltonian over some interaction time $t$.

The wave function $\phi (x)$ of the pointer is assumed to be
\begin{equation}
\phi (x)=\langle x|\phi\rangle =\Big({1\over \sqrt{2\pi}\sigma}\Big)^{1/2}\exp\Big( -{(x-x_0)^2\over 4\sigma^2}\Big).
\label{eq:wave}
\end{equation}
Here, we have introduced the centre $x_0\ne 0$, which is essential in the discussion of 2.4.  The initial system-pointer state $|\Phi(0)\rangle=|I\rangle |\phi\rangle$ evolves obeying
\begin{equation}
i\hbar{d\over dt}|\Phi(t)\rangle =\hat H|\Phi (t)\rangle \sim\hat H_I|\Phi (t)\rangle ,
\end{equation}
to 
\begin{equation}
|\Phi (t)\rangle =|I\rangle |\phi\rangle -{igt\over\hbar}\hat A|I\rangle\hat\pi |\phi\rangle +\cdots ,\label{eq:gtichiji}
\end{equation}
where $|I\rangle$ is the initial state of the observed system.  The state of the pointer $|\phi_{ji}\rangle$ both after the interaction between the observed system and the probe and the post-selection in $|\psi_j\rangle$ is, up to the lowest order in $gt$, 
\begin{equation}
|\phi_{ji}\rangle\equiv\frac{\langle\psi_j|\Phi (t)\rangle}{\langle\psi_j|I\rangle} = |\phi\rangle -{igt\over\hbar}\frac{\langle\psi_j|\hat A|I\rangle}{\langle\psi_j|I\rangle}                                                                                                                                               \hat\pi |\phi\rangle .\label{eq:phifi}
\end{equation}
Then, the expectation value of the pointer's position $\hat x$ for this state is
\begin{equation}
\langle\phi_{ji}|\hat x|\phi_{ji}\rangle
=x_0+gt\Re\langle\hat A\rangle_{\psi_j,I},\label{eq:wmwv}
\end{equation}
where 
\begin{equation}
\langle\hat A\rangle_{\psi_j,I}\equiv\frac{\langle\psi_j |\hat A|I\rangle}{\langle\psi_j |I\rangle}\label{eq:teigi}
\end{equation}
is the weak value of an operator $\hat A$ for an initial state $|I\rangle$ and a final state $|\psi_j\rangle$.

\subsection{Ordinary interpretation of weak values}

In some papers\cite{Aha3}\cite{Hosoya2}, it is considered as a basis of statistical interpretation of weak values that ordinary expectation values can be described as the sum of corresponding weak values.   We examine the expectation value $\langle I |\hat A |I\rangle$ of an observable $\hat A$ for a state vector $|I \rangle$.  Let $|\psi_j\rangle$ be the eigenvectors that correspond to the respective eigenvalues $\psi_j,\ j=1,2,\cdots$ of an observable $\hat \Psi$.   By assuming that a set of projection operators $\{|\psi_j\rangle\langle\psi_j|\}$ are complete, i.e., $1=\sum_j|\psi_j\rangle\langle\psi_j|$ and that $\langle\psi_j|I\rangle\ne0$,  
\begin{equation}
\begin{split}
\langle I |\hat A |I\rangle &=\sum_j\langle I |\psi_j\rangle\langle\psi_j |\hat A |I\rangle \\
&=\sum_j{\rm Pr}(\psi_j|I)\langle\hat A\rangle_{\psi_j,I},\label{eq:0}
\end{split}
\end{equation}
where
\[
{\rm Pr}(\psi_j|I)=|\langle I |\psi_j\rangle |^2
\]
is the probability that the state $|\psi_j\rangle$ is found in the state $|I\rangle$.  Thus, we can interpret the expectation value $\langle I |\hat A |I\rangle$ as a statistical average of the weak values $\langle\hat A\rangle_{\psi_j,I}$, and as a result, weak values are treated by many authors as the expectation values of $\hat A$ between the initial state $|I\rangle$ and the final states $|\psi_j\rangle,\ j=1,2,\cdots$.  However, as shown below, we should not decide based exclusively on (\ref{eq:0})  whether weak values can be interpreted as probabilities or expectation values.

We write the proposition `an eigenvalue $a_i$ is obtained when an observable $\hat A$ is measured' as $A(a_i)$, and its corresponding projection operator is denoted $\hat A_i=|a_i\rangle\langle a_i|$.  Similarly, we define a proposition $\Psi(\psi_j)$ and a projection operator $\hat\Psi_j=|\psi_j\rangle\langle\psi_j|$.  A set of such propositions constitutes a $\sigma$-complete orthomodular lattice\cite{Svozil}\cite{Maeda}, as does the corresponding set of such projection operators.

Let $\hat A$ in (\ref{eq:0}) be the projection operator $ \hat A_i=|a_i\rangle\langle a_i |$.  Then,
\begin{equation}
\langle I |\psi_j\rangle \langle \psi_j |a_i\rangle\langle a_i |I\rangle ={\rm Pr}(\psi_j|I )\langle\hat A_i\rangle_{\psi_j,I}.\label{eq:2}
\end{equation}
A necessary and sufficient condition for the operator $\hat \Psi_j\hat A_i$ to be a projection operator is $[\hat \Psi_j,\hat A_i]=0$.  If and only if this condition is satisfied, $\hat \Psi_j\hat A_i$ corresponds to a proposition $\Psi(\psi_j)\wedge A(a_i)$ and the left-hand side of (\ref{eq:2}) is its probability for $|I\rangle$\cite{Maeda}.  Here, we define the {\it joint probability} of $A(a_i)$ and $\Psi (\psi_j)$ for $|I\rangle$ as the probability of the proposition $\Psi(\psi_j)\wedge A(a_i)$, i.e., the probability of finding $|\psi_j\rangle$ {\it and} $|a_i\rangle$ in $|I\rangle$ {\it simultaneously}.  We do not regard the probability of finding $A(a_i)$ and then $\Psi(\psi_j)$ {\it one by one} as the joint probability, because the operation of $\hat A_i$ must affect the probability of $\Psi(\psi_j)$ for $|I\rangle$, as shown in the subsection {\it 2.3}.  Thus, the weak value $\langle\hat A\rangle_{\psi_j,I}$ is the conditional probability of finding $|a_i\rangle$ in $|I\rangle$ when $|\psi_j\rangle$ is found in $|I\rangle$ if and only if $[\hat \Psi_j,\hat A_i]=0$.  Then,
\[
0\le\langle I |\psi_j\rangle \langle \psi_j |a_i\rangle\langle a_i |I\rangle\le{\rm Pr}(\psi_j|I )\le 1,
\]
and hence,
\begin{equation}
0\le\langle\hat A_i\rangle_{\psi_j,\Phi}\le 1.
\end{equation}
As shown later, the weak values are actually 0 or 1 in such a case.  
We can interchange $|\psi_j\rangle$ and $|I\rangle$ in the above discussion.  If $|I\rangle\langle I |$ and $\hat\Psi_j$ commute,  $\langle\hat A\rangle_{\psi_j,I}$ is the probability of finding $|a_i\rangle$ in $|I\rangle$ (or in $|\psi_j\rangle$).  

If $[\hat \Psi_j,\hat A]\ne 0$, the projection operator that corresponds to a proposition $\Psi(\psi_j)\wedge A(a_i)$ is $\lim_{n\rightarrow \infty}(\hat\Psi_j \hat A_i)^n$\cite{Svozil}.  Instead, if we construct (for example) a Hermitian operator $\hat A\hat\Psi\hat A$ and a projection operator $ |h_k\rangle\langle h_k|$, where $\hat A\hat\Psi\hat A |h_k\rangle =h_k |h_k\rangle$, then the proposition corresponding to $A\Psi A(h_k)$ exists.  Nevertheless, this proposition is not expressed with the help of the $\Psi (\psi_j)$s and/or $A(a_i)$s.  In contrast, either $\hat \Psi_j\hat A_i$ is not a projection operator or it does not correspond to any propositions.  Thus, if any two of $\hat\Psi_j$, $\hat A_i$ and $|I\rangle\langle I |$ do not commute, we cannot interpret the left-hand side of (\ref{eq:2}) as a probability or the right-hand side of (\ref{eq:0}) as a sum of probabilities.   Therefore, in such cases,  $\langle\hat A_i\rangle_{\psi_j,I}$ is not the conditional probability of finding $|a_i\rangle$ in $|I\rangle$ when $|\psi_j\rangle$ is found in $|I\rangle$,

To clarify the meaning of strange {\it probability}, we divide $\langle I |\hat \Psi_j\hat A_i|I \rangle$ into its real part and imaginary part as follows:
\begin{equation}\begin{split}
\langle I |\hat \Psi_j\hat A_i|I \rangle=&\langle I |{1\over 2}(\hat \Psi_j\hat A_i+ \hat A_i\hat \Psi_j)|I \rangle \\
&+\langle I |{1\over 2} [\hat\Psi_j ,\hat A_i] |I\rangle.
\end{split}
\end{equation}
Thus, the weak value
\begin{equation}
\langle \hat A\rangle_{\psi_j,I}={\langle I |\hat\Psi_j\hat A_i |I\rangle\over{\rm Pr}(\psi_j|I)}\label{eq:wv}
\end{equation}
becomes real if $\langle I |[\hat \Psi_j,\hat A_i]|I\rangle= 0$.  Here, we should pay attention to the fact that even if $\langle |[\hat \Psi_j,\hat A_i]|\rangle= 0$ for some states, this is not a sufficient condition for $[\hat \Psi_j,\hat A_i]=0$, i.e., this condition does not ensure that  
$\hat \Psi_j\hat A_i$ is a projection operator and possesses the corresponding proposition.   If  $\langle I |[\hat \Psi_j,\hat A_i]|I\rangle= 0$ and any pair of $\hat\Psi_j$, $\hat A_i$ and $|I\rangle\langle I |$ do not commute, (\ref{eq:wv}) may be more than 1 or less than 0.  This possibility is not strange because (\ref{eq:wv}) is not a (conditional) probability as shown above.  Considering Hardy's paradox, we will encounter such a situation.  

We corroborate the above conclusion by reexamining $\langle I |\hat A|I\rangle$.  When $\hat A=\hat A_i$,
\begin{equation}
\begin{split}
\langle I |\hat A_i|I\rangle&=\langle I |\hat A_i\hat A_i|I\rangle\\
&=\sum_j\langle I |\hat A_i\hat \Psi_j\hat A_i|I\rangle\\
&=\sum_j{\rm Pr}(\psi_j|I)|\langle\hat A_i\rangle_{\psi_j,I}|^2.\label{eq:0dash}
\end{split}\end{equation}
$\langle I|\hat A_i\hat\Psi_j\hat A_i|I\rangle =\langle I|\hat \Psi_j\hat A_i|I\rangle$ if  $\hat A_i$ and $\hat \Psi_j$ commute.  Then, by comparing (\ref{eq:0dash}) and (\ref{eq:0}), it is clear that $\langle\hat A_i\rangle_{\psi_j,I}=0$ or $1$.  Conversely, if $\langle\hat A_i\rangle_{\psi_j,I}\ne |\langle\hat A_i\rangle_{\psi_j,I} |^2$, it is obvious that at least one of the following two statements is false: `$\langle\hat A_i\rangle_{\psi_j,I}$ is the expectation value of $\hat A_i$ between an initial state $|I\rangle$ and a final state $|\psi_j\rangle$' or `$|\langle\hat A_i\rangle_{\psi_j,I}|^2$ is the expectation value of $\hat A_i$ between an initial state $|I\rangle$ and a final state $|\psi_j\rangle$'.  We have demonstrated above that the former statement is false if the operators do not commute, and we will demonstrate below that the latter statement is also false if they do not commute.

 As written by Aharonov et al.\cite{Aha1},
\begin{equation}
|\langle\hat A_i\rangle_{\psi_j,I} |^2 = {{\rm Pr}(a_i|\psi_j){\rm Pr}(a_i|I)\over {\rm Pr}(\psi_j|I)}.\label{eq:jijou}
\end{equation}
Because the denominator of the right-hand side does not depend on $a_i$, $|\langle\hat A_i\rangle_{\psi_j,I} |^2$ gives the product of two independent probabilities {\rm Pr}($a_i |\psi_j$) and {\rm Pr}($a_i |I$) (divided by Pr$(\psi_j|I )$).  It is worth noting that (\ref{eq:jijou}) is not a conditional probability if $[\hat \Psi_j,\hat A_i]\ne0$.  To verify this fact, we rewrite (\ref{eq:jijou}) as
 \begin{equation}
|\langle\hat A_i\rangle_{\psi_j,I} |^2 = {\langle I |\hat A_i\hat \Psi_j\hat A_i |I\rangle \over {\rm Pr}(\psi_j|I)}.\label{eq:jijou2}
\end{equation}
The right-hand side of this equation is the expectation value of {\it one} observable $\hat A_i\hat\Psi_j\hat A_i$ divided by Pr$(\psi_j|I )$.  If $[\hat \Psi_j,\hat A_i]\ne0$, then $\hat A_i\hat\Psi_j\hat A_i$ corresponds to no proposition, and consequently,  (\ref{eq:jijou}) is not a conditional probability because $\hat A_i\hat\Psi_j\hat A_i$ is not a projection operator.  

The above discussion can be straightforwardly applied to other observables, such as $\hat A=\sum_ia_i\hat A_i$.   Thus, it is obvious that if $\hat \Psi_j$ and $\hat A$ do not commute, then the weak value $\langle\hat A\rangle_{\psi_j,I}$ is not the conditional expectation value of $\hat A$ for $|I\rangle$ when $|\psi_j\rangle$ is found in $|I\rangle$.

\subsection{POVM of weak measurement}

As shown in the previous subsection, we can not regard (\ref{eq:0}) as a basis of the statistical interpretation of weak values.  However, some authors\cite{Wise2}\cite{Dre2}\cite{Dressel}\cite{Dre3}\cite{Hof1}\cite{Hof2} have developed discussions with the help of {\it positive operator-valued measure} (POVM).  Let $\{\hat M_m\}$ be a set of operators that act on the Hilbert space of the observed system.  The probability of obtaining an outcome $m$ for the quantum state expressed in a density matrix $\hat \rho$, Pr$(m|\rho)$, is
\begin{equation}
\Pr (m|\rho)={\rm Tr} \Big\{\hat M_m\hat \rho\hat M^{\dagger}_m\Big\}={\rm Tr} \Big\{\hat M^{\dagger}_m\hat M_m\hat \rho\Big\},
\end{equation}
where $\{\hat M^{\dagger}_m\hat M_m\}$ is POVM that satisfies
\begin{equation}
\sum_m\hat M^{\dagger}_m\hat M_m=1.
\end{equation}

Next, let us consider a sequential measurement corresponding to two sets of POVMs, $\{\hat M^{(1)\dagger}_m\hat M^{(1)}_m\}$ and $\{\hat M^{(2)\dagger}_n\hat M^{(2)}_n\}$.  Then, the probability of obtaining the first outcome $m$ and the second outcome $n$, $\Pr (n,m|\rho)$, is
\begin{equation}
\Pr (n,m|\rho)={\rm Tr}\Big\{\hat M^{(2)}_n\hat M^{(1)}_m\hat\rho\hat M^{(1)\dagger}_m\hat M^{(2)\dagger}_n\Big\}.
\end{equation}
Dressel {\it et al.}\cite{Dre2} and Wiseman\cite{Wise2} have considered
\begin{equation}
\frac{\Pr (n,m|\rho )}{\sum_m\Pr (n,m|\rho )}\label{eq:rennzoku}
\end{equation}
as a conditional probability or a probability between some initial state and final state.   Some of the authors  have insisted that (\ref{eq:rennzoku}) would become the corresponding weak value with $\hat \rho =|I\rangle\langle I|$, $\hat M^{(1)\dagger}_m\hat M^{(1)}_m=\hat A_m$, $\hat M^{(2)\dagger}_n\hat M^{(2)}_n=|\psi_n\rangle\langle\psi_n|$  in its weak coupling limit, and hence, the weak value could be interpreted as a conditional probability.  However, if so, (\ref{eq:rennzoku}) could take negative values despite the fact that it is made up of the sum, product  and quotient of some probabilities.  This inconsistency is not the matter of interpretation.  It is the matter of calculation.

To raise the point, we examine their calculation\cite{Dre2}.  Let $\hat A$ be 
\begin{equation}
\hat A=\sum_m\hat A_m=\sum_m\alpha_m\hat M^{(1)\dagger}_m\hat M^{(1)}_m\label{eq:MM}.
\end{equation}
They have defined the {\it conditional expectation value} $_n\langle\hat A\rangle$ as the expectation value obtained by the sequential measurement,
\begin{equation}
_n\langle\hat A\rangle=\frac{\sum_m\alpha_m\Pr (n,m|\rho )}{\sum_m\Pr (n,m|\rho )}.\label{eq:heikinnchi}
\end{equation}
The POVM $\hat E^{(1)}_m$ that corresponds to $\hat M^{(1)}_m$ is expanded up to the lowest order of $g$, the constant that gives the strength of the measurement:

\begin{equation}
\hat E^{(1)}_m=\hat M^{(1)\dagger}_m\hat M^{(1)}_m=p_m\hat 1+g\hat E^{(1)\prime}_m,\label{eq:povm}
\end{equation}
where
\[
\sum_mp_m=1.
\]
Substituting (\ref{eq:povm}) into (\ref{eq:heikinnchi}) leads to
\begin{equation}
_n\langle\hat A\rangle=\frac{{\rm Tr}\big[\hat M^{(2)\dagger}_n\hat M^{(2)}_n\{\hat A, \hat\rho\}\big]}{2{\rm Tr}\big[\hat M^{(2)\dagger}_n\hat M^{(2)}_n\hat\rho\big]}.                                      
\end{equation}
In addition, if $\hat \rho=|I\rangle\langle I|$, $\hat M^{(2)\dagger}_n\hat M^{(2)}_n=|\psi_n\rangle\langle\psi_n|$, then
\begin{equation}
_n\langle\hat A\rangle=\Re\frac{\langle\psi_n|\hat A|I\rangle}{\langle\psi_n|I\rangle}.
\end{equation}
Here, we should pay attention to the fact that the right-hand side of this equation is not the real part of the weak value in the meaning defined in (\ref{eq:teigi}) .  Rather,
\begin{equation}
\begin{split}
_n\langle\hat A\rangle&=\sum_m\alpha_mp_m+g\Re\langle\hat A^{\prime}\rangle_{\psi_n,I}\\
&=\  _n\langle\hat A\rangle\big|_{g=0}+g\Re\langle\hat A^{\prime}\rangle_{\psi_n,I},\label{eq:inevitable}
\end{split}
\end{equation}
where
\[
\hat A^{\prime}\equiv \sum_m\alpha_m\hat E^{(1)\prime}_m.
\]
Taking account of the fact that (\ref{eq:heikinnchi}) is an expectation value obtained by sequential measurement, the real part of the weak value, $\Re\langle\hat A^{\prime}\rangle_{\psi_n,I}$, can be regarded as the indicator of the inevitable back-action, caused by the weak measurement, to the post-selection, i.e., to the measurement of $|\psi_n\rangle\langle\psi_n|$. Here, the inevitable back-action is defined as $\langle I|\hat A\hat\Psi_j\hat A|I\rangle -\langle I|\hat\Psi_j|I\rangle$ for the strong measurement.   Moreover, we notice that the term {\it weak measurement} means the interaction described by the Hamiltonian (\ref{eq:Hamiltonian}) - in other words, the POVM (\ref{eq:povm}).

However, Hofmann\cite{Hof1} has insisted that the back-action of a weak measurement to the post-selection should be the second order of  the coupling constant.  He has calculated the expectation values of the operators corresponding to the post-selection for the state after the weak measurement with the help of POVM, and he has demonstrated that their sum should not contain the back-action of the weak measurement up to the first order of the coupling constant.  Although the result of his calculation is right, what has been demonstrated is that the first order of the back-action should vanish in the {\it sum}.  The back-action to {\it each} post-selection has not been calculated in \cite{Hof1}.

\subsection{Weak value as the indicator of back-action}

To clarify the above conclusion, let us reconsider the operation of weak measurement and post-selection.  In post-selection, a final state $|\psi_j\rangle$ is selected after the weak measurement of an operator $\hat A$.  In other words, the final state is obtained as the state projected by measuring the expectation value of the operator $|\psi_j\rangle\langle\psi_j|$ after the weak measurement.

From (\ref{eq:gtichiji}), let us define $|\Phi (t)_{\phi}\rangle$ as the state both after the measurement of the position of the pointer and before the post-selection, i.e.,
\begin{equation}
\begin{split}
|\Phi (t)_{\phi}\rangle&\equiv x_0^{\ -1}\langle\phi |\hat x|\Phi (t)\rangle\\
& = |I\rangle -{igt\over\hbar x_0}\langle\phi |\hat x\hat\pi|\phi\rangle\hat A|I\rangle.\label{eq:position}
\end{split}
\end{equation}
The expectation value of the operator $|\psi_j\rangle\langle\psi_j|$ for this state is, up to the first order of $gt$,
\begin{equation}
\langle\Phi (t)_{\phi}|\psi_j\rangle\langle\psi_j|\Phi (t)_{\phi}\rangle=
|\langle \psi_j|I\rangle |^2+{gt\over x_0}\Re \big(\langle I|\psi_j\rangle\langle\psi_j|\hat A|I\rangle\big).
\end{equation}
Thus,
\begin{equation}
\Re\frac{\langle\psi_j|\hat A|I\rangle}{\langle\psi_j|I\rangle}=\frac{x_0}{gt}\frac{|\langle\psi_j|\Phi (t)_{\phi}\rangle |^2-|\langle \psi_j|I\rangle |^2}{|\langle \psi_j|I\rangle |^2}.\label{eq:jitsubu}
\end{equation}
With the help of (\ref{eq:wmwv}), this equation is rewritten as 
\begin{equation}
\frac{\langle\phi_{fi}|\hat x|\phi_{fi}\rangle}{x_0}=\frac{|\langle\psi_j|\Phi (t)_{\phi}\rangle |^2}{|\langle \psi_j|I\rangle |^2}.\label{eq:sokuteichi}
\end{equation}
Then, (\ref{eq:jitsubu}) and (\ref{eq:sokuteichi}) are the operational expression of (\ref{eq:inevitable}).  These are our main results.  Although it has already been noted in \cite{Ste1}\cite{Ste2}\cite{Jozsa}\cite{Dressel} that the imaginary part of a weak value gives the back-action caused by the weak measurement, its real part is interpreted as a conditional probability or a conditional expectation value there.  Nevertheless, the above equations demonstrate that the real part of the weak value and the expectation value of the position of the pointer after the post-selection give only the back-action caused by the weak measurement.  It is worth noting that this back-action itself does not depend on the probe system and is inevitable, as shown in the previous subsection, especially (\ref{eq:inevitable}).    Thus, we have no reason to interpret (\ref{eq:jitsubu}) as a conditional probability or a conditional expectation value, if any pair of the operators $|I\rangle\langle I|$, $|\psi_j\rangle\langle\psi_j|$ and  $\hat A$ do not commute.

It is easier to convince ourselves of this fact, if we consider the following concepts.  After weak measurement, only a small part of the entangled state of the observed system and the probe would change, whereas almost the whole state would remain the same as the initial state.  As noted in \cite{Bliokh}, a weak measurement of {\it one} particle and that of {\it many} identical particles should give the same result, if the measurement is repeated many times and the average is adopted.  
Thus, we can suppose that the initial state of a weak measurement is formed of many identical particles.  When this state is weakly measured, a small number of particles are strongly measured and the rest are not measured, though which one is measured is never determined.  The weak value demonstrates how the strong measurement for the minority changes the initial state.

Next, let us consider a complete set of projection operators, 
$\{\hat A_i\}$, which satisfy
\[
\hat 1=\sum_i\hat A_i.
\]
As easily confirmed from the definition (7),
\begin{equation}
\sum_i\langle\hat A_i\rangle_{\psi_j,I}=1.\label{eq:complete}
\end{equation}
(\ref{eq:complete}) is often treated as supporting evidence of the statistical interpretation of weak values.  We reconsider ({\ref{eq:complete}) based on the above discussion.  The interaction Hamiltonian between a projection operator $\hat A_i$ and the corresponding probe is
\begin{equation}
\hat H_{I(i)}\equiv g\hat A_i\hat\pi_i,
\end{equation}
where $\hat\pi_i$ is the momentum of the pointer.  We assume that the wave function $\phi_i(x_i)$ of each pointer is in the same form as (\ref{eq:wave}).  $|\Psi (t)_i\rangle$ is defined like (\ref{eq:position}) as the state both after the measurement of the position of the {\it i}-th pointer and before the post-selection:
\begin{equation}
|\Psi (t)_i\rangle= |I\rangle -\frac{igt}{\hbar x_0}\langle\phi_i|\hat x_i\hat\pi_i|\phi_i\rangle\hat A_i|I\rangle.
\end{equation}
Similarly, the state after the measurement of the positions of all the pointers corresponding to their respective projection operators of $\{\hat A_i\}$ is defined as
\begin{equation}
\begin{split}
|\Phi (t)_1\rangle&= |I\rangle -\frac{igt}{\hbar x_0}\sum_i\langle\phi_i|\hat x_i\hat\pi_i|\phi_i\rangle\hat A_i|I\rangle\\
&=\Big(1-\frac{igt}{\hbar x_0}\langle\phi |\hat x\hat\pi |\phi\rangle\Big)|I\rangle.
\end{split}
\end{equation}
It is  in the nature of things that this state is identified with the initial state except for the normalisation factor.  Then, 
\begin{equation}
\sum_i\Re\frac{\langle\psi_j|\hat A_i|I\rangle}{\langle\psi_j|I\rangle}=\frac{x_0}{gt}\frac{|\langle\psi_j|\Phi (t)_{1}\rangle |^2-|\langle\psi_j|I\rangle |^2}{|\langle\psi_i|I\rangle |^2}=1.\label{eq:jitsubu1}
\end{equation}
This equation demonstrates that (\ref{eq:complete}) only reflects the fact that the operation of the identity operator does not affect the state.

\section{Hardy's paradox}
Recently, the counter factual statements of Hardy's paradox\cite{Hardy} were interpreted with the help of weak values\cite{Aha4} and it was ascertained that they agreed with the values obtained by the corresponding weak measurement\cite{Irvine}\cite{Lundeen}\cite{Yokota}. 
We investigate the weak values in Hardy's paradox based on the discussion in the previous section.

As shown in Fig.1, a device composed of an electron Mach-Zehnder interferrometer (MZI$^-$) and a similar machine with positrons (MZI$^+$) is examined.  OL is the domain where these two MZIs overlap.  We assume that pair annihilation must occur if an electron(e$^-$) and a positron(e$^+$) exist simultaneously in OL.  The length between BS1$^{-(+)}$ and BS2$^{-(+)}$ is adjusted to let e$^-$ (e$^+$) be detected by a detector C$^{-(+)}$ without exception in a solo MZI$^{-(+)}$ experiment.  Conversely, detection by a detector D$^{-(+)}$ implies that  obstacles exist on either path.  

\includegraphics{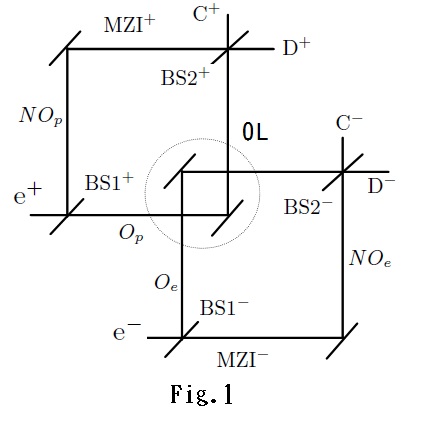}

We consider the case where the pair annihilation does not occur and e$^{-}$ and e$^+$ are detected by D$^-$ and D$^+$, respectively.  The initial state $|\Phi\rangle$ and the final state $|\Psi\rangle$ are  defined as
\begin{equation}\begin{split}
|\Phi\rangle = {1\over\sqrt{3}}\Big[ &|O_p,NO_e\rangle + |NO_p,O_e\rangle\\
& + |NO_p,NO_e\rangle\Big],
\end{split}
\end{equation}
\begin{equation}\begin{split}
|\Psi\rangle =& {1\over2}\Big[ |O_p,O_e\rangle - |O_p,NO_e\rangle\\
& - |NO_p,O_e\rangle + |NO_p,NO_e\rangle\Big],
\end{split}
\end{equation}
where $O$ and $NO$ are abbreviations of `{\it Through OL}' and `{\it Not through OL}', respectively.  
 Then $\big|\langle\Psi |\Phi\rangle\big|^2  = {1\over 12}$ by ordinary quantum mechanical calculation. However, the weak values are
\begin{equation}
\langle \hat N^{+,-}_{O,O}\rangle_{\Psi, \Phi} = 0,\label{eq:oo}
\end{equation}
\begin{equation}
\langle \hat N^{+,-}_{O,NO}\rangle_{\Psi, \Phi} =\langle \hat N^{+,-}_{NO,O}\rangle_{\Psi, \Phi} = 1,
\end{equation}
\begin{equation}
\langle \hat N^{+,-}_{NO,NO}\rangle_{\Psi, \Phi} = -1,\label{eq:nono}
\end{equation}
\begin{equation}
\langle \hat N^{\pm}_O\rangle_{\Psi, \Phi} = 1,\label{eq:ohitotsu}
\end{equation}
\begin{equation}
\langle \hat N^{\pm}_{NO}\rangle_{\Psi, \Phi} = 0,\label{eq:nohitotsu}
\end{equation}
where
\begin{equation}\begin{split}
\hat N^{+,-}_{O(NO),O(NO)}=|&O(NO)_p,O(NO)_e\rangle\\
\otimes&\langle O(NO)_p,O(NO)_e|,\label{eq:number2}\end{split}
\end{equation}
\begin{equation}
\hat N^{+}_{O(NO)}=\hat N^{+,-}_{O(NO),O}+\hat N^{+,-}_{O(NO),NO}, \label{eq:number1}
\end{equation}
\begin{equation}
\hat N^{-}_{O(NO)}=\hat N^{+,-}_{O,O(NO)}+\hat N^{+,-}_{NO,O(NO)}. \label{eq:number3}
\end{equation}

It is easily verified that any two of $\hat \Psi\equiv|\Psi\rangle\langle\Psi |$, $\hat\Phi\equiv|\Phi\rangle\langle\Phi |$ and any one of the operators defined in (\ref{eq:number2}) - (\ref{eq:number3}) do not commute.  For example, though $\langle\Phi|[\hat \Psi ,\hat N^{+,-}_{NO,NO}]|\Phi\rangle=\langle\Psi|[\hat \Phi ,\hat N^{+,-}_{NO,NO}]|\Psi\rangle =0$,
\begin{equation}\begin{split}
\hat\Psi\hat N^{+,-}_{NO,NO}\hat \Psi\hat N^{+,-}_{NO,NO}&={1\over 4}\hat \Psi\hat N^{+,-}_{NO,NO},\\
\hat\Phi\hat N^{+,-}_{NO,NO}\hat \Phi\hat N^{+,-}_{NO,NO}&={1\over 3}\hat \Phi\hat N^{+,-}_{NO,NO}.\\ 
\label{eq:zero}\end{split}
\end{equation}
Therefore, as shown in the previous section, $\langle \hat N^{+,-}_{NO,NO}\rangle_{\Psi, \Phi}$ cannot be regarded as the  conditional probability of finding both e$^-$ and e$^+$ on NOs between the initial state $|\Phi\rangle$ and the final state $|\Psi\rangle$. The discussions of $\hat N^{+,-}_{O,NO}$, $\hat N^{+,-}_{NO,O}$ and $\hat N^{+,-}_{O,O}$ are roughly equivalent.  Thus, regardless of which operator is used, even with the help of the weak values (\ref{eq:oo}) - (\ref{eq:nohitotsu}), we have no right to evaluate the validity of the counterfactual statement `e$^-$ must pass through OL to ensure that e$^+$ is detected by D$^+$ and vice versa.  Nevertheless, both e$^-$ and e$^+$ cannot simultaneously pass through OL,  because they must be annihilated together if they encounter each other'. 
 We can discuss the three-box paradox\cite{Aha6}\cite{Vaidman}\cite{Resch} similarly.

\section{W-slit experiment}

As is well known, the interference pattern is lost if we attempt to determine the photon's trajectory in the double-slit experiment.  However, Kocsis {\it et al.}\cite{Kocsis} have experimentally define a set of trajectories for an ensemble of the photons that pass through a double slit with the help of weak measurement technique.  They have used the polarisation degree of freedom of the photons as the pointer that weakly couples to their momenta.  After the weak measurement of the momenta, they have selected the subensemble of the photons arriving at a particular position by the strong measurement.  Thus, they insisted that the average momentum of the photons reaching any particular position in the image plane could be determined, and their average trajectories could be reconstructed by repeating this procedure in a series of planes.

In their study, the average momentum of the photon's subensemble post-selected at a position $\xi$ should be given in the form
\begin{equation}
\langle\hat P\rangle_{\xi,I}={\langle I |\xi\rangle\langle \xi|\hat P|I\rangle\over\big |\langle \xi|I\rangle\big |^2},\label{eq:Pwv}
\end{equation}
where $\hat P$ is the momentum operator of $\xi$-direction.  However, we cannot interpret the weak value (\ref{eq:Pwv}) as a conditional expectation value of $\hat P$ between the initial state $|I\rangle$ and the final state $|\xi\rangle$, because $|\xi\rangle\langle\xi |$ and $\hat P$ do not commute.  That is, we cannot interpret  (\ref{eq:Pwv}) as the average momentum of the photons that have reached the position $\xi$.  On the other hand, if we replace $\hat A$ with $\hat P$ and $|\psi_j\rangle$ with $|\xi\rangle$ in (\ref{eq:jitsubu}), 
\begin{equation}
\Re\langle\hat P\rangle_{\xi,I}=\frac{x_0}{gt}\frac{|\langle \xi|\Phi (t)_{\phi}\rangle |^2-|\langle \xi|I\rangle |^2}{|\langle \xi|I\rangle |^2},\label{eq:kakuritsu}
\end{equation}
where
\begin{equation}
|\Phi (t)_{\phi}\rangle =|I\rangle -\frac{igt}{\hbar x_0}\langle\phi |\hat x\hat\pi |\phi\rangle\hat P|I\rangle.
\end{equation}
This equation indicates that $\Re\langle\hat P\rangle_{\xi,I}$ gives the back-action caused by the weak measurement of the momentum to the post-selection of the position.  In addition, it is obvious from (\ref{eq:kakuritsu}) that the real part of (\ref{eq:Pwv}) is a classically measurable quantity, as noted in \cite{Bliokh}.  

In \cite{Wise1}, which is one of the bases of \cite{Kocsis}, the average velocity of a particle at a position $\xi$, $v(\xi ,t)$, is operationally defined as 
\begin{equation}
v(\xi ,t)\equiv\lim_{\tau\rightarrow 0}\tau^{-1}{\rm E}\big[\xi_{strong}(t+\tau)-\xi_{weak}(t)\big|\xi_{strong}=\xi\big]\label{eq:ref2},
\end{equation}
where $\xi_{strong(weak)}(t)$ is the strongly (weakly) measured position of the particle at the time $t$ and ${\rm E}\big[a\big|F\big]$ is the average of $a$ when $F$ is true.  If the Hamiltonian of the observed system is $P^2/2m + V(\xi)$, (\ref{eq:ref2}) is identified with Bohm's velocity\cite{Bohm} and in proportion to $\Re\langle\hat P\rangle_{\xi,I}$.  However, we know from the above discussion that ${\rm E}\big[\xi_{weak}(t)\big|\xi_{strong}=\xi\big]$ is not the average position at the time $t$ when the position $\xi$ is post-selected at the time $t+\tau$, and hence, we cannot interpret (\ref{eq:ref2}) as the average velocity.  Therefore, the fact that $\Re\langle\hat P\rangle_{\xi,I}$ is in proportion to Bohm's velocity does not help the claim in \cite{Kocsis}.

\section{Conclusion}

The main conclusions of this study, which were drawn in the second section, are as follows:  The real part of the weak value $\langle\hat A\rangle_{\psi_j,I}$ is not the expectation value of an operator $\hat A$ between an initial state $|I\rangle$ and a final state $|\psi_j\rangle$.  Rather, it gives the back-action caused by the weak measurement of $\hat A$ for $|I\rangle$, which changes the probability of finding the state $|\psi_j\rangle$.  There are so many studies based on the statistical interpretation of weak values that we cannot examine all of them.  However, the studies investigated in the previous two sections are typical, and we have noted some of their essential faults.  Therefore, we conclude that it is worth considering suggestion that there is controversy in previous findings, though our ideas may not be applicable to all discussions concerning weak values and weak measurement.

\end{document}